\documentclass[twocolumn,prl]{revtex4}
\usepackage{graphicx}

\begin{document}


  \title{Boundary-mediated electron-electron interactions in quantum point contacts}

\author{V.~T.~Renard$^{1,2}$,
O.~A.~Tkachenko$^{2,3}$, V.~A.~Tkachenko$^{3}$, T. Ota$^{1}$, N.
Kumada$^{1}$, J-C Portal$^{2,4}$, Y. Hirayama$^{1,5,6}$. }

\affiliation{$^1$ NTT Basic Research Laboratories, NTT Corporation
3-1 Morinosato Wakamiya, Atsugi 243-0198, Japan} \affiliation{$^2$
GHMFL-CNRS, BP 166, F-38042, Grenoble Cedex 9,
France,}\affiliation{$^3$ Institute of Semiconductor Physics,
630090 Novosibirsk, Russia,}\affiliation{$^{4}$~INSA and Institut
Universitaire de France, F-31077, Toulouse Cedex 4, France.}
\affiliation{$^5$ SORST-JST 4-1-8 Honmachi, Kawaguchi, Saitama
331-0012, Japan} \affiliation{$^6$ Department of Physics, Tohoku
University, Sendai 980-8578, Japan}

\date{March 28, 2008}

\begin{abstract}
An unusual increase of the conductance with temperature is
observed in clean quantum point contacts for conductances larger
than $2e^2/h$. At the same time a positive magnetoresistance
arises at high temperatures.  A model accounting for
electron-electron interactions mediated by boundaries (scattering
on Friedel oscillations) qualitatively describes the observation.
It is supported by numerical simulation at zero magnetic field.
\end{abstract}


\maketitle


Quantum point contacts (QPC) are usually formed when two wide 2D
conducting regions are connected by a small constriction. They
exhibit a number of intriguing phenomena among which conductance
quantization \cite{Wees88,Wharam88} is the most emblematic.
Nowadays QPCs are very common tools for condensed matter
physicists. Examples of recent applications include nuclear spin
manipulation, solid state electron optics or precise electron
counting \cite{Applications}. 
Recently, the puzzling low conductance ``0.7
anomaly"\cite{Thomas96} has attracted most of the attention since
it is related to electron-electron (e-e) interactions. But apart
``0.7 anomaly" it is commonly believed that the physics of QPCs is
well understood using a single-particle picture (See
Ref.\onlinecite{Beenakker91} and references therein). On the
contrary, it is well known that the properties of two-dimensional
electron gases (2DEGs) dramatically depend on interactions
\cite{Altshuler85,Aleiner97,Zala01,Gornyi,Renard05}. The related
corrections to the conductance and tunnelling density of states
have their origin in Friedel oscillations (FO) of electron density
around impurities. 
Friedel oscillations are also known to appear at boundaries
\cite{Alekseev98,Aleiner98,Sablikov} and could therefore affect
the properties of nano-scale devices. Interestingly, boundary
mediated FOs were recently shown to be possibly involved in "0.7
anomaly" physics \cite{Rejec06} but this subject is still highly
debated \cite{Zozoulenko07}.\\In this context we show that e-e
interactions can have a significant influence on transport
properties of clean quantum point contacts even in the large
conductance regime ($G>>2e^2/h$). We used a combination of
relatively low electron densities, high mobility and low series
resistance to clearly uncover the effects of e-e interactions. It
increases interactions due to reduced screening and ensures that
impurity scattering can be disregarded. Starting from the second
conductance step, the average slope of the conductance versus gate
voltage linearly increases with increasing temperature $T$. At the
same time, the low field magneto-resistance (MR) is non-monotonic
at high $T$ and strongly temperature dependent. Some of our
findings are present, although not as evident, in previous works
\cite{Hansen02,Pyshkin00,Cronenwett02,Feng99}. They are consistent
with a model of e-e interaction mediated by boundaries which is
supported by numerical simulation at $B=0$ T.

The quantum point contacts were defined employing a split-gates
lateral depletion technique \cite{Thornton86} on high mobility
GaAs quantum wells \cite{note}. Needle and square shaped samples
of various sizes were used to check the influence of the geometry
(see Table \ref{table1}). All samples a have back gate to tune the
electron density and some samples have a $0.2$ $\mu$m-wide center
gate to obtain better defined conductance quantization steps when
grounded \cite{Lee06}. Four-terminal resistance measurements were
carried out between $350$ mK and $10$ K with a standard
low-frequency technique at small excitation voltage $<20$ $\mu$V
to avoid heating effects. Any obvious temperature dependance
(activated parallel conduction, anomalous temperature dependance
of the series 2DEG or leakage of the gates) was excluded
\cite{note1}. More than ten samples showed a qualitatively similar
behavior.

\begin{table}[htbp]
\caption{\label{table1}Summary of the samples (for the wafer and
Hall bar see \cite{note}).}
\begin{ruledtabular}
\begin{tabular}{ccccc}
Sample&Wafer&Hall bar&point contact&center gate\\
\hline
Square1 & 1 & 1 & W=0.6 $\mu$m; L=0.4 $\mu$m&yes\\      
Square2 & 1 & 1 & W=0.6 $\mu$m; L=1 $\mu$m&yes\\        
Needle1 & 2 & 2 & W=0.8 $\mu$m&no\\                     
Needle2 & 2 & 2 & W=0.6 $\mu$m&no\\                     

\end{tabular}
\end{ruledtabular}
\end{table}

\begin{figure}[htbp]
\includegraphics[angle=0,width=0.9\columnwidth]{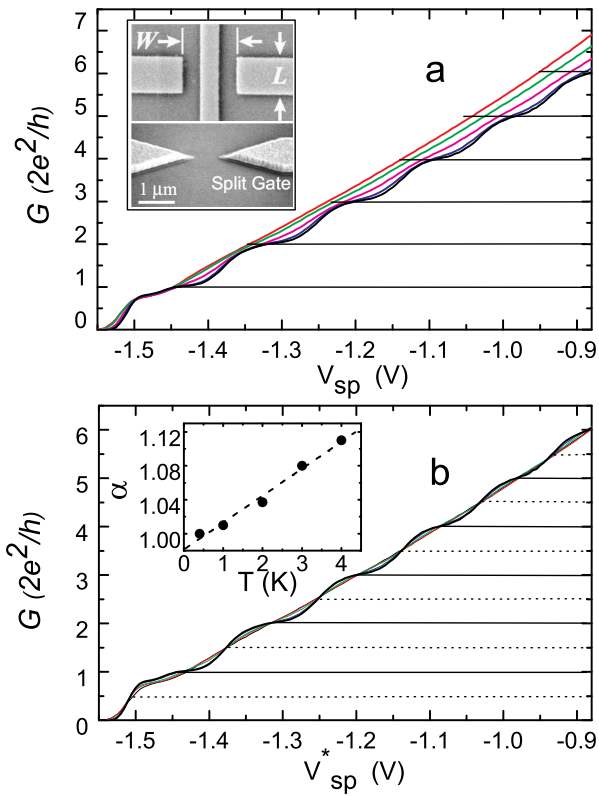}%
\caption{\label{Figure1} a) Conductance of the sample Square1 as a
function of $V_{sp}$ (From top to bottom $T$=4 K, 3 K, 2 K, 1 K and
350 mK). The inset shows SEM images illustrating possible shapes
of the point contacts. b) Conductance as a function of the scaled
split-gate voltage $V_{sp}^*$ for the same temperatures. The inset
shows the scaling parameter $\alpha$.}
\end{figure}

Figure \ref{Figure1}a shows the conductance $G$ of the sample
Square1 as the split-gate voltage $V_{sp}$ is varied for different
temperatures. The small ($\sim 15 \Omega$) 2DEG series resistance
was not subtracted. At low $T$ the conductance is quantized to
exact integer values of $2\times e^2/h$. Note that the ``0.7
anomaly" is observed. Increasing temperature not only shrinks the
plateaus but also increases the overall slope of $G(V_{sp})$
resulting in the increase of conductance with $T$. Such a change
in the slope of the conductance is not expected from a simple
energy averaging, which produces fix temperature independent points
in the conductance located at integers of $e^2/h$ \cite{Wees91}.
Regions with alternatively positive or negative temperature
dependence should be separated by these fixed points. This is not
observed in our experiment. It is however possible to restore this
behavior by a phenomenological approach. For each temperature a new
effective split-gate voltage can be calculated
$V_{sp}^*=V_{p}+\alpha(T)\times(V_{sp}-V_{p})$ to superpose the
curve to the low temperature one (See Fig. \ref{Figure1}b). Here,
$V_p$ is the pinch off voltage and $\alpha$ is the scaling
parameter. The strong linear temperature dependence of $\alpha$
points at quantum effects, possibly e-e interactions.

In the Hartee-Fock approximation scattering on the potential
created by all the other electrons can be considered as the origin
of e-e interactions effects
\cite{Altshuler85,Aleiner97,Zala01,Gornyi,Aleiner98,Sablikov}.
This potential is connected to the formation of Friedel
oscillations of the electron density close to scatterers (defects
or boundaries) due to interferences between incoming and reflected
electron waves. Spatial variations of their density results in a
varying effective potential seen by conducting electrons. In our
case the QPC is clean and FOs are created at the boundaries of the
constriction. They affect its conductance in two different ways.
At low $T$ when the thermal length $l_T$ exceeds  $W$ the QPC
width, FOs located in the 2DEG can reduce the conductance at a
plateau via scattering of emitted electrons back to the QPC
\cite{Alekseev98}. As $T$ is increased these oscillations are
damped and conductance plateaus reach ideal integer value of
$2\times e^2/h$ \cite{Yacoby96}. When $l_T\approx W$ there remain
only FOs located inside the QPC. This corresponds to our
experimental conditions ($l_T\approx$ 1 $\mu$m at $T=300$ mK).
As $T$ is further increased the transverse oscillations inside the
QPC are damped thus effectively widening the constriction (or
equivalently shifting 1D energy levels). One has to compensate by
applying a lower $V_{sp}$ to obtain a comparable value of
conductance (the curves shift to the left). The scaling can be
qualitatively understood as follows. Electrostatically defined
QPCs have a parabolic cross section potential
$U=m^*\omega^2x^2/2$, $m^*$ is the effective mass. The width of
the channel can be defined as
 $W=(2/\omega)\sqrt{2E_F/m^*}$, where $E_F$
is the Fermi level in 2DEG reservoirs. One-dimensional energy levels
are written $E_n=(n+1/2) \hbar\omega$ and in a constant
capacitance model $W$ is proportional to $V_{sp}-V_p$. It follows
that $\Delta E_n/E_n \propto \Delta \omega/\omega \propto \Delta
W/W \propto \Delta (V_{sp}-V_p)/(V_{sp}-V_p)$. In analogy to the
2D case where the conductivity $\sigma$ is renormalized as
$\Delta\sigma/\sigma \sim \beta\times T/E_F$ in the ballistic
regime of interaction \cite{Zala01} we expect that $\Delta E_n/
E_n\sim \beta\times T/E_F$ when $E_n \sim E_F$ (the
interaction-induced shift $\Delta E_n$ of the closest to Fermi
level subband only is considered). Here, $\beta$ is the
interaction parameter. This leads to the rescaling $V_{sp}\to
V_{sp}^*=V_p+\alpha(T)(V_{sp}-V_p)$ with $\alpha(T) \propto T$ as
measured.

\begin{figure*}[htbp]
\parbox{1.3\columnwidth}{\includegraphics[width=1.3\columnwidth]{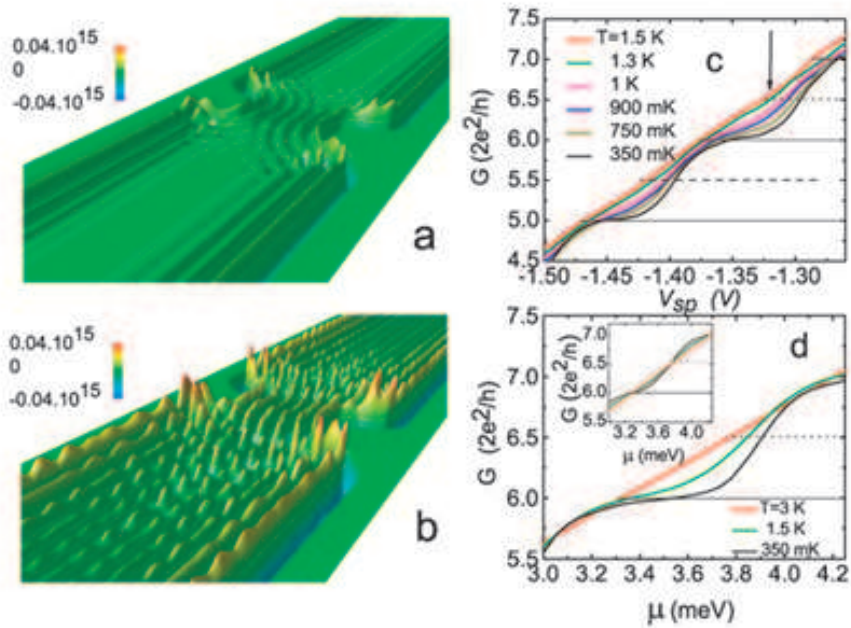}}%
\parbox{0.05\columnwidth}{~}
\parbox{0.6\columnwidth}{\caption{\label{Figure2} Calculated correction to the electron
density due to Friedel oscillations at $G\approx6\times 2e^2/h$
for T=3.5 K (a) and T=350 mK (b). The scale is m$^{-2}$. This
clearly illustrate the rise of the interaction potential as $T$
decreases. c) Conductance of the sample Needle1 for different
temperature. The vertical arrow corresponds to the voltage at
which Fig. \ref{Figure3}a was recorded. d) Calculated conductance
at a similar value of conductance. The inset shows the calculated
conductance neglecting e-e interactions.} }
\end{figure*}

This result is consistant with a numerical non-self-consistent
Hartree-Fock simulation. The principle of the computation is first
to calculate the local electron density $n(x,y,T,\mu)$ solving the
one-particle Shr\"{o}dinger equation in a ``soft wall"
electrostatic potential for different chemical potential $\mu$ and
temperature $T$. Indeed, instead of calculating the conductance as
a function of $V_{sp}$ for a varying potential $U_0(x,y,V_{sp})$
and fixed $\mu$, we chose to fix the electrostatic potential and
vary the chemical potential. This greatly simplifies the
calculation because the wave-functions are calculated only once.
It is valid for small interval of split gate voltage (between two
conductance steps). Numerical technique and parameters are
described in Ref. \onlinecite{note2,Usuki95}.
 In analogy to the 2D case, the interaction
correction to the potential is computed as follows $\delta
U(x,y,T,\mu)=\beta\delta n(x,y,T,T^*,\mu)/D$, where $\beta$ is the
interaction parameter, $D$ is the 2D density of states, $\delta
n(x,y,T,\mu)=n(x,y,T,\mu)-n(x,y,T^*,\mu)$ and $T^*$ a temperature
high enough to suppress FOs. The obtained correction which
includes only $T$-dependent terms is then used to solve the
transmission problem in the total potential which is now a
function of $T$ and $\mu$ \cite{note2}. Contrary to 2D systems
containing impurities \cite{Zala01,Gornyi} it is not possible to
relate the interaction parameter $\beta$ to the interaction
constant $F_0$ in a simple way. Close to the boundaries the
electron density drops to zero. This drop is not completely abrupt
and the first Friedel oscillation develops in rarefied
weakly-screening electron gas where e-e interactions become
particularly strong. This is taken into account in our simulation
by the large value of the parameter $\beta=-6$. Figures
\ref{Figure2}a and b show the obtained correction to the electrons
density $\delta n(x,y)$ in the vicinity of the constriction at
$G\approx 6\times e^2/h$ for $T$=3.5 K (a) and $T$=350 mK (b). This
clearly illustrates the formation of FOs as $T$ is decreased. Note
that pronounced features are demonstrated despite the use of a ``soft wall" model.\\
Figures \ref{Figure2}c and d compare the experimental
conductance of the sample Needle1 to the conductance calculated by
the method described above. In both cases the fix temperature
independent points are missing and the conductance is on average
increasing with temperature. A test of the model consists in
neglecting the interaction term of the scattering potential (i.e.
setting the interaction parameter $\beta$ to 0) which restores the
dependence for energy averaging as expected (Inset to Fig.
\ref{Figure2}d). An overall good qualitative agreement is obtained
between computer and real experiments which further confirms the
qualitative understanding.

\begin{figure}[htbp]
\includegraphics[angle=0,width=0.85\columnwidth]{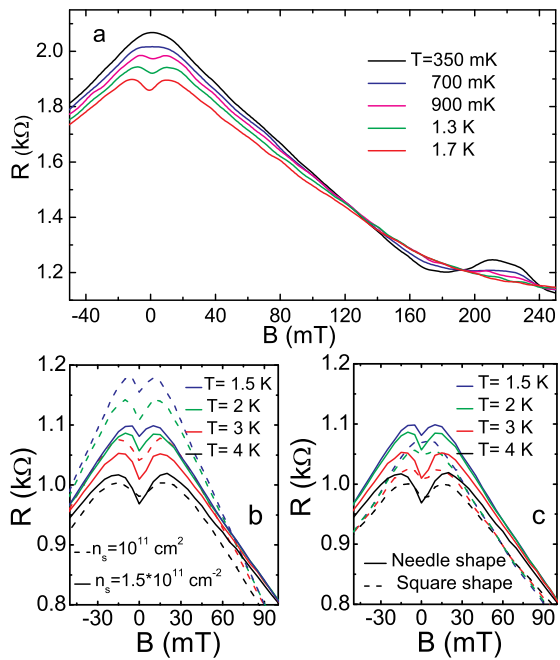}
\caption{\label{Figure3} a) Resistance of the sample Needle1 at
the gate voltage indicated by an arrow in Fig. \ref{Figure2}c. b)
Resistance of the sample Needle2 for the electron density $n_s=1\times10^{11}$
cm$^{-2}$ (dash) and $n_s=1.5\times10^{11}$ cm$^{-2}$ (solid) for
$G(T=4 $K$)\approx 13\times2e^2/h$. c) Resistance of Needle2
(solid) and Square2 (dot) for $n_s=1.5\times10^{11}$ cm$^{-2}$ and
$G(T=4 $K$)\approx 13\times2e^2/h$.}
\end{figure}

In general, the magnetoresistance reveals important informations
about scattering, coherent processes and e-e interactions
\cite{Gornyi}. Figure~\ref{Figure3}a shows that beside the unusual
temperature dependence at $B= 0$ T the samples present an
unexpected field dependence. The data displayed were measured for
sample Needle1 at $G\approx6\times2e^2$/h (see arrow in Fig.
\ref{Figure2}c). At high temperature, the magneto-resistance
presents a maximum around $B=10$ mT and becomes negative at higher
fields. At low $T$ the resistance decreases with $B$ at all
magnetic fields. The high field slope of the MR is $T$-dependent.
Note that the maximum moves to lower fields in QPCs with larger
width (not shown).

A linear negative MR with a slope inversely proportional to the
width of the channel is known to appear in QPCs
\cite{VanHouten88}. The increased slope at low temperature that we
observe is consistent with the observation at $B$ = 0 T which lead
to the conclusion that e-e interactions could in principle reduce
the effective width of the channel at low $T$. As for the positive
MR, it cannot be attributed to diffuse boundary scattering which
has very different characteristics \cite{Boundary}. In particular
it is $T$-independent and absent in short constrictions.
Similarly, the interplay of boundary scattering and electron
collisions \cite{Gurzhy95} can be discarded. Finally,
commensurability of electron trajectories with the voltage probes
can be ruled out since very different Hall bar geometries were
tested (Table. \ref{table1}).

The absence of positive MR at low $T$ confirms that the observed
quantum effect is due to e-e interaction and is not an
interference effect (i.e. Weak anti-localization) which should
increase at low $T$. Similarly to 2D systems \cite{Zala02}, it can
most likely be explained by a variation of the parameter $\beta$
at small magnetic fields which is only visible at high $T$. Note
that simulations with $B$-independent $\beta$ did not produce the
positive MR. Figure \ref{Figure3}b and c demonstrates that the
magneto-resistance depends on the electron density and detailed
geometry of the point contacts. Carefully adjusting the back- and
split-gate voltages, the conductance $G(T=4$K$,B=0 $T$)$ of the
sample Needle2 was tuned to the same value
($\approx13\times2e^2/h$) for two different electron densities.
Although qualitatively similar, the effect is found to be more
pronounced at the lower density. This is consistent with the
Friedel oscillation picture which should have in principle smaller
effect at high densities when the system resembles a
non-interacting Fermi liquid. Figure \ref{Figure3}c compares the
results of the sample Needle2 to the effect obtained in the sample
Square2 at the same density and conductance. The temperature
dependence at $B=0$ T is comparable but the field dependence is
much more pronounced for the sample Needle2 which has smoother
entrances. The geometry dependence therefore appears to be a very
interesting tool to study the effect of magnetic field on the
measured effect. The positive MR raises interesting questions
about the influence of low magnetic field on Friedel oscillations
and requires additional theoretical work.

We believe that the presented deviations have not been observed in
regular point contacts (for example in Ref. \cite{Wees91}) due to
the lower mobility and higher density in these experiments.
Indeed, Friedel oscillations are known to depend exponentially on
disorder and interaction. However, recent measurements on similar
samples show similar phenomenology (see Ref.
\onlinecite{Pyshkin00,Hansen02,Cronenwett02} for the conductance
quantization and Ref. \onlinecite{Feng99} for the MR) indicating
that our observation is a general effect. It could have particular
importance in nano-scale electronics since boundaries dominate
transport properties in nano-devices (electrostatically defined quantum dots, rings, Y-junctions etc.).

\begin{acknowledgments}
We are very grateful to I. Gornyi, A. Dmitriev, K. Takashina, Y.
Tokura, K. Pyshkin for helpful discussions and K. Muraki for
providing some samples. O.A.T. acknowledges UJF for the invitation
as ``Ma\^{i}tre de Conf\'{e}rences", the Supercomputing Siberian
center and CNRS/IDRIS (project 61778). Part of this work was
supported by CNRS/RAS agreement between ISP Novosibirsk and GHMFL
Grenoble.
\end{acknowledgments}



\begin{thebibliography}{99}
\bibitem{note}
    Two wafers were used with the mobility $\mu_1\approx4\times10^6$ cm$^2$/Vs and $\mu_2\approx3\times10^6$ cm$^2$/Vs for a density
    $n_{s1}=1.5\times10^{11}$ and $n_{s2}=1\times10^{11}$. They were processed into 10 $\mu$m wide (type 1) and 100 $\mu$m wide (type 2) Hall bars.
\bibitem{note1}
    Hall measurements show that the electron density is $T$-independent. The resistance of the 2DEG decreases only by a few Ohms between high and low $T$ which
    makes this source of $T$-dependence incompatible with the observation. Also, the split gate's leakage current  was
    monitored to be smaller than 20 pA at all voltages and T.
\bibitem{note2}
    The computation region was 3 $\mu$m long and 0.7 $\mu$m wide. The constriction is 250 nm wide in the model. The calculation step along x and y was  5 nm, the energy step was $10^{-3}$ meV.
    Usual hopping contant for GaAs were used.
\end{thebibliography}
\end{document}